\documentstyle[12pt]{article}
\def\a{\alpha}\def\b{\beta}\def\g{\gamma}
\def\k{\kappa}
\def\Th{\Theta}\def\Om{\Omega}\def\G{\Gamma}
\def\y{\vartheta}

\newcommand{\nn}{\nonumber\\}\newcommand{\p}[1]{(\ref{#1})}
\begin{document}
\renewcommand{\thefootnote}{\arabic{footnote}}
\setcounter{footnote}0
\begin{center}
{\Large \bf A generalized action principle for D=4 doubly supersymmetric
membrane}\\

\vspace{0.5cm}

{\large\bf Aleksey Yu. Nurmagambetov} \\
NSC Kharkov Institute of Physics and Technology\\
Akademicheskaya 1\\
310108 Kharkov, Ukraine\\
e-mail: kfti@rocket.kharkov.ua
\end{center}

\begin{abstract}
In this review I discuss some recent results concerning D=4 doubly
supersymmetric membranes within the framework of geometrical
approach obtained in the collaboration with Igor Bandos, Dmitrij Sorokin
and Dmitrij Volkov.
\end{abstract}

Starting its development from the beginning of the 70's the string theory
has taken up a stable place into the modern theoretical physics and
nowadays it can be considered as one of the most appropriate candidates
for the construction of the Unified Theory. One of many at the moment, but
not the only one, because if we have made a very important step from point
to string why don't we go further towards a consideration of membranes and
other extended objects? In the other words we need find an answer
whether the string theory is an unique possibility for unification of all
interactions or it is not so.

The first steps of membrane physics having in some sence purely academic
value were more than unassuming in comparison with the impressive
achievements of the (super)string theory. But discovering D=11
supergravity which is a low energy limit of D=11 {\sl
supermembranes} arroused an interest to the relativistic
supermembranes as an alternative to the superstring theory.

The absence of Green -- Schwarz superstring covariant quantization
procedure and impossibility to construct a perturbative picture of string
interactions were ones of many reasons which stimulated the
investigations of the string/membrane duality \cite{duff} at the low~--~
energy effective action level in the framework of non -- perturbative
approach. These investigations have legalized a status of higher --
dimensional extended objects (super~--~p~--~branes) still more and incited
the search of a version of super~--~p~--~branes description being
suitable for covariant quantization.

In contrast to the superstrings the problem of super~--~p~--~brane
covariant quantization is hampered first of all due to the more
complicated structure of these objects, giving essentially nonlinear
equations of motion, as well as by other obstacles, containing,
particularly, an infinite reducuble $\k$ -- like symmetry. Nowadays there
are no recipes to solve completely membrane equations of motion even
at a pure bosonic level, although the recent papers of J. Hoppe et al.
\cite{hoppe} inspire a hope for a positive reply to this question.
Unfortunately, there is also no idea how to quantize gauge
theories in the presence of infinite reducible symmetries.

In spite of difficulties mentioned above, at least the $\k$ -- symmetry
problem doesn't seem to be so dangerous now. The progress in solution for
this problem is closely related to the development of twistor -- like
\cite{stv} and Lorentz -- harmonic \cite{bz} approaches to the doubly
supersymmetric particles and strings.

In the twistor -- like approach the local fermionic $\k$ -- symmetry,
generated by infinite reducible constraint, is replaced by local
worldsheet supersymmetry which is irreducible by definition.

However, clear understanding of geometrical nature of the $\k$ -- symmetry
in the framework of doubly supersymmetric twistor -- like approach is not
the whole story because the general structure of superparticle and
superstring action has the following form (see for details \cite{gs}):
\begin{equation}\label{1}
S=\int\,d^{p}\xi\,d^{N}\eta\,P_{K}\Pi^{K}+\ \ generalization\ \ of\ \ WZ\
\ term
\end{equation}
with $P_{K}$ are the superfield Lagrange multipliers and superfield
functions $\Pi^{K}$ containing all the geometrical constraints of
supersurface embedding into a target superspace and equations of motion.
The geometrical and physical sence of superfield Lagrange multipliers is
shadowed, moreover this theory possesses another infinite
reducible symmetry \cite{gs} which eliminates auxiliary degrees of
freedom contained in the superfield Lagrange multipliers.

The previous attempts to avoid the difficulties pointed above were not
realized completely in the case of D=10 superparticle \cite{bnsv} but have
led to the construction of pure geometrical picture of
super~--~p~--~branes supersurface embedding \cite{bpstv} and suggestion of
a generalized action principle for super~--~p~--~branes \cite{bsv}.

The geometrical approach to super~--~p~--~branes supported by the idea
of supergravity rheonomic theory gives many advantages in comparison with
known formulations. For this reason the detailed programm of investigations
of extended relativistic objects in the framework of this formalism can be
outlined.

The first step in this direction was made by Igor Bandos, Dmitrij Sorokin
and Dmitrij Volkov \cite{bsv} where a generalized action
principle for super -- p -- branes was proposed.

One of the next steps of this approach \cite{igor}, applied to the D=4
supermembranes is shortly presented in this report.

Let us start from the following action for D=4
closed supermembrane \cite{bsv}:
\begin{equation}\label{2}
S_{ D=4,2} =-{1\over{2}}
\int_{{\cal M}_{3}}\left(E^{a_0}
e^{a_1}e^{a_{2}}\varepsilon_{a_0a_1a_{2}}
-{2\over{3}}
e^{a_0} e^{a_1}e^{a_{2}}\varepsilon_{a_0a_1a_{2}}\right) \nn
\pm{4\over{3}}
\int_{{\cal M}_{3}}
\Pi^{\underline{m_2}}\Pi^{\underline{m_1}}
\Th\G_{\underline{m}_1\underline{m}_2}d\Th,
\end{equation}

The main features of this action is that it is constructed out of
differential forms being target and world supersurface
vielbeins without any application of Hodge operation and is integrated
over the {\it bosonic submanifold} ${\cal M}_{3}$ of the whole world
supersurface; $\varepsilon_{a_0a_1a_{2}}$ is the unit antisymmetric tensor
and membrane tension is chosen to be one.

So, for the construction of this action we choose a basis of one forms on a
world supersurface
\begin{equation}\label{3}
e^{A}=(e^{a},e^{\a})\ \ \ ; \ \ \ e^{A}={e^{A}}_{M}dz^{M}
\end{equation}
and the latter is parameterized by the coordinates
$z^{M}=\{\xi^{a},\eta^{\a}\}$.

An arbitrary local frame in the flat target superspace can be obtained
from the pullback onto world supersurface of the basic supercovariant
forms \cite{dva}
\begin{equation}\label{4}
\Pi^{\underline{m}}=dX^{\underline{m}}-id\Th\sigma^{\underline{m}}\bar{\Th}
+i\Th\sigma^{\underline{m}}d\bar{\Th},
\ \ \ \ d\Th^{\underline{\mu}},\ \ \ \ d\bar{\Th}^{\underline{\dot{\mu}}}
\end{equation}
by $SO(1,D-1)$ rotations
\begin{equation}\label{5}
E^{\underline{a}}=\Pi^{\underline{m}}u_{\underline{m}}^{\underline{a}},
\ \ \
E^{\underline{\a}}=d\Th^{\underline{\mu}}\y_{\underline{\mu}}^
{\underline{\a}},\ \ \ \bar{E}^{\underline{\dot\a}}=
d\bar{\Th}^{\underline{\dot\mu}}\bar{\y}_{\underline{\dot\mu}}^
{\underline{\dot\a}}
\end{equation}
with the vector $u_{\underline{m}}^{\underline{a}}$ and spinor $(
\y_{\underline{\mu}}^{\underline{\a}},\bar{\y}_{\underline{\dot\mu}}^
{\underline{\dot\a}})$ components of the local Lorentz frame
(supervielbein) in the target superspace, belonging to the $SO(1,D-1)$ and
doubly covering for the latter $Spin(1,D-1)$ groups respectively \cite{bz}
(our index notations and conventions are closely related to the ones in
ref.  \cite{bsv}).

The superfield equations of motion derived by variation over
$u_{\underline{m}}^{a}$ and $e^{a}$ variables
$$\Pi^{\underline{m}}u_{\underline{m}}^{\perp}e^{a_{1}}e^{a_{2}}
\varepsilon_{a_{0}a_{1}a_{2}}=0$$
\begin{equation}\label{6}
(\Pi^{\underline{m}}u_{\underline{m}}^{a_{0}}-e^{a_{0}})e^{a_{1}}
\varepsilon_{a_{0}a_{1}a_{2}}=0
\end{equation}
lead to a part of the rheotropic conditions
\begin{equation}\label{7}
E^{a}\equiv\Pi^{\underline{m}}u_{\underline{m}}^{a}=e^{a}
\end{equation}
which relate the vector target space
vielbeins to the world supersurface ones and being the standard relations
of geometrical approach \cite{barab} which can be put by hands. In the
framework of our consideration these relations are obtained from the
variational principle and mean that $e^{a}$ is induced by embedding.

Having in mind the rheotropic conditions \p{7} and expression for pullback
vector one~--~form
\begin{equation}\label{8}
\Pi^{\underline{m}}=e^{\a}\Pi_{\a}^{\underline{m}}+e^{a}\Pi_{a}^
{\underline{m}}
\end{equation}
we obtain the "geometrodynamical" condition \cite{gs} which was used
previously as one specifying embedding world supersurface into a target
space
\begin{equation}\label{9}
\Pi_{\a}^{\underline{m}}=D_{\a}X^{\underline{m}}-iD_{\a}\Th\sigma^
{\underline{m}}\bar{\Th}+i\Th\sigma^{\underline{m}}D_{\a}\bar{\Th}=0
\end{equation}

The integrability condition for \p{9}
\begin{equation}\label{10}
{\g^{a}}_{\a\b}\Pi_{a}^{\underline{m}}=D_{\a}\Th\sigma^{\underline{m}}
D_{\b}\bar{\Th}
\end{equation}
looks like vector -- spinor relations
$$u_a^{\underline{\a}\underline{\dot\a}}=
(\g_a)^{\a\b}
\y_\a^{\underline{\a}}{\bar\y}_\b^{\underline{\dot\a}}$$
and we can restrict some components of target space basic one -- forms to
the following values
\footnote{To prove the rought equality we need to take into account an
analog of Weyl symmetry \p{19} and fix an appropriate gauge.}
\begin{equation}\label{11}
\Pi^{\underline{m}}_{a}\sim{u^{\underline{m}}_{a}};\ \ \ \
D_{\a}\Th^{\underline{\b}}\sim{\y^{\underline{\b}}_{\a}}
\end{equation}
After such a choice the appearence of Virasoro -- like
constraint
\begin{equation}\label{12}
\Pi^{\underline{m}}_{a}\Pi_{\underline{m}b}=\eta_{ab}
\end{equation}
in addition to the "geometrodynamical" condition \p{9} becomes perfectly
clear.

But on the other hand the integrability condition for \p{8}
\begin{equation}\label{13}
d\Pi^{\underline{m}}=-2id\Th\sigma^{\underline{m}}d\bar{\Th}=
T^{a}u^{\underline{m}}_{a}+e^{a}Du^{\underline{m}}_{a}
\end{equation}
involves the world supersurface torsion
\begin{equation}\label{14}
T^{a}=De^{a}\equiv{de^{a}-{\Om^{a}}_{b}e^{b}}
\end{equation}
with the $\Om^{ab}(d)={1\over2}u_{\underline{m}}^{a}du^{\underline{m}b}$
being the $SO(1,2)$ connection induced by the embedding, i.e. \cite{bpstv}
\begin{equation}\label{15}
\Om^{ab}(D)=0
\end{equation}

The requirement \p{11} demands some restrictions on the
world supersurface torsion. In particular
\footnote{This result completely coincides with integrability condition
for the rheotropic relations \p{7}.}
\begin{equation}\label{16}
T^{a}=-2id\Th\sigma^{\underline{m}}d\bar{\Th}u_{\underline{m}}^{a}
\end{equation}
and, consequently,
$$T^{a}_{\a\b}=-2i(\g^{a})_{\a\b}$$
\begin{equation}\label{17}
T^{a}_{\a{b}}=0;\ \ \ \ T^{a}_{cb}=-i\chi_{c}\g^{a}\chi_{b}
\end{equation}
where $2Im\chi_{c}^{\a}\equiv{D_{c}\Th^{\underline{\a}}\y_{\underline{\a}}^
{\a}-D_{c}\bar{\Th}^{\underline{\dot{\a}}}\bar{\y}_{\underline{\dot{\a}}}^
{\a}}$ is the matter superfield defined below
\footnote{First constraint of \p{17} is a backbone for all the SYM and
supergravity superfield formulations giving a possibility to extract the
physical sector of the theory.}.

Equations of motion derived from the action \p{2} by the variation over
$\Th^{\underline{\a}}$ and $\bar{\Th}^{\underline{\dot{\a}}}$ variables
have the form of:
\begin{equation}\label{18}
d\bar{\Th}^{\underline{\dot{\a}}}u^{a_{0}}_{\underline{\a}\underline{
\dot{\a}}}e^{a_{1}}e^{a_{2}}\varepsilon_{a_{0}a_{1}a_{2}}\pm{
\varepsilon_{\underline{\dot{\a}}\underline{\dot{\b}}}{\Pi^{\underline{
\dot{\a}}}}_{\underline{\a}}\Pi^{\underline{\dot{\b}}\underline{\b}}
d\Th_{\underline{\b}}}=0
\end{equation}
$$+h.c.$$

Analysis of \p{18} with taking into account a part of the rheotropic
conditions \p{7} and an analog of Weyl
symmetry
$$e^{a}\rightarrow{\hat{e}^{a}=W^{2}e^{a}}$$
\begin{equation}\label{19}
e^{\a}\rightarrow{\hat{e}^{\a}=We^{\a}-{1\over2}ie^{b}(\g_{b})^{\a\b}D_{\g}W}
\end{equation}
leads to the remaining part of the rheotrotic relations (we fix the gauge
$W=-1$)
$$
E^{\a}\equiv{d\Th^{\underline{\a}}\y_{\underline{
\a}}^{\a}=e^{\a}+ie^{a}\chi_{a}^{\a}}$$
\begin{equation}\label{20}
\bar{E}^{\a}\equiv{d\bar{\Th}^{\underline{\dot{\a}}}\bar{\y}_{\underline{
\dot{\a}}}^{\a}=e^{\a}-ie^{a}\chi_{a}^{\a}}
\end{equation}
together with supermembrane equation of motion:
\begin{equation}\label{21}
{(\g^{a})_{\a}}^{\b}\chi_{a}^{\a}=0
\end{equation}

The integrability conditions for spinor part of target superspace vielbeins
\begin{equation}\label{22}
dd\Th^{\underline{\a}}=0;\ \ \ \ dd\bar{\Th}^{\underline{\dot{\a}}}=0
\end{equation}
which are coincided with integrability conditions for \p{20}, give another
world supersurface torsion constraints:
\begin{equation}\label{23}
T^{\b}=-{i\over2}e^{a}\chi_{a}^{\a}{(\g_{c})_{\a}}^{\b}\Om^{c\perp}
\Rightarrow{T^{\b}_{\a\g}=0}
\end{equation}
expression for $SO(1,3)/SO(1,2)$ vielbein form $\Om^{\perp}_{\a\b}\sim{
(\g^{a})_{\a\b}\Om^{\perp}_{a}}$ \cite{bpstv} in terms of $\chi_{a}^{\g}$
\begin{equation}\label{23}
\Om^{\perp}_{\a\b}=-2ie^{\g}\chi_{\underline{\a\b\g}}-e^{b}D_{\{\a\vert}
\chi_{b\vert\b\}}
\end{equation}
and restriction on the $\chi_{a}^{\g}$ superfield:
\begin{equation}\label{24}
D_{[c}\chi^{\b}_{a]}=2i(\chi_{a}\g^{s}\chi_{c})\chi_{s}^{\b}
\end{equation}
Here $\underline{\a\b\g}$, $\{\dots\}$ and $[\dots]$ denote the complete
symmetrization, symmetrization and antisymmetrization respectively.

Moreover, the integrability conditions \p{22} lead to the equation of
motion \p{21} remaining only spin $3/2$ non -- trivial part of $\chi^{\a}_
{a}$
\begin{equation}\label{25}
\chi_{\a\b\g}\equiv{\g^{a}_{\a\b}\chi_{a\g}=\chi_{\underline{\a\b\g}}}
\end{equation}

Geometry of world supersurface is completely described in terms of
$SO(1,2)$ connection $\Om^{ab}(d)$ and $SO(1,3)/SO(1,2)$ vielbein
$\Om^{a\perp}={1\over2}u_{\underline{m}}^{a}du^{\underline{m}\perp}$
\cite{barab} up to rotation and translation of world supersurface at whole.

By definition $\Om^{ab}$ and $\Om^{a\perp}$ satisfy Maurer -- Cartan
equations in the form of Peterson~--~Codazzi equation
\begin{equation}\label{26}
d\Om^{a\perp}(d)-{\Om^{a}}_{b}(d)\Om^{b\perp}(d)=0
\end{equation}
and Gauss equation
\begin{equation}\label{27}
R^{ab}(d,d)=d\Om^{ab}(d)-{\Om^{a}}_{c}(d)\Om^{cb}(d)=\Om^{a\perp}(d)
\Om^{b\perp}(d)
\end{equation}

Intrinsic and induced geometry of world supersurface defined by the
$SO(1,2)$ connection, $SO(1,3)/SO(1,2)$ vielbein and pullback of a
target space basic one -- forms respectively is coincided due to
expression \p{15} \cite{bpstv}.

The independent part of Peterson -- Codazzi equation gives another
restrictions on the $\chi^{\a}_{a}$ superfield and Gauss equation allows
to us to get an expression for supersurface curvature $R_{\a\b}\sim{
(\g^{ab})_{\a\b}R_{ab}}$ in terms of Rarita -- Schwinger -- like
superfield:
\begin{equation}\label{28}
R_{\a\b}=-4e^{\g}e^{\delta}\varepsilon^{\varepsilon\xi}\chi_{\underline{
\a\g\xi}}\chi_{\underline{\delta\varepsilon\b}}+4ie^{b}e^{\delta}D^{\g}
\chi_{b\{\a\vert}\chi_{{\underline{\g\delta\b}\}}}+e^{b}e^{c}D_{\a}\chi^{\g}
_{b}D_{\g}\chi_{c\b}
\end{equation}

In the framework of our approach world supersurface supergravity Bianchi
identities
\begin{equation}\label{29}
DT^{A}={1\over2}e^{B}{R_{B}}^{A}
\end{equation}
are fulfilled automatically and equations of motion for $X$ variables are
dependent from the other ones.

Thus we have a complete description of our superfield membrane formulation
in terms of world supersurface supergravity and Rarita -- Schwinger --
like matter superfield.

Unfortunately, our formulation does not possess "off -- shell" invariance:
the rheotropic relations lead to the equations of motion and,
consequently, there is no any possibility to construct an alternative
superfield formulation a l\'a Galperin and Sokatchev \cite{gs} as it can be
done in the cases of D=3,10 heterotic superstring.

\vspace{0.5cm}
{\bf Acknowledgements}

I would like to thank the organizers of the Xth International Workshop on
HEP\&QFT where this report was presented for the possibility of my
participation in this meeting and for the kind hospitality in Zvenigorod.

I'm very grateful to Dmitrij Volkov and Igor Bandos for fruitful
discussions of a generalized action principle conception and
usefull remarks and Dmitrij Sorokin for valued discussions and remarks.

This work was supported in part by the ISF grant RY9200 and by
the INTAS grant 93-493.

\end{document}